\documentclass[conference]{IEEEtran}
\IEEEoverridecommandlockouts
\usepackage{cite}
\usepackage{amsmath,amssymb,amsfonts}
\usepackage{algorithmic}
\usepackage{graphicx}
\usepackage{textcomp}
\usepackage{xcolor}
\usepackage{booktabs}
\usepackage{multirow}
\usepackage{subfigure}
\def\BibTeX{{\rm B\kern-.05em{\sc i\kern-.025em b}\kern-.08em
    T\kern-.1667em\lower.7ex\hbox{E}\kern-.125emX}}
\begin{document}

\title{Transformer and GAN Based Super-Resolution Reconstruction Network for Medical Images\\
{\footnotesize \textsuperscript{*}}
}

\author{\IEEEauthorblockN{1\textsuperscript{st} Weizhi Du}
\IEEEauthorblockA{\textit{The Athenian School} \\
Danville, CA \\
23wdu@athenian.org}
\and
\IEEEauthorblockN{2\textsuperscript{nd} Shihao Tian}
\IEEEauthorblockA{\textit{Department of Electric and Computing Engineering} \\
\textit{Cornell University}\\
Ithaca, NY \\
st689@cornell.edu}
}

\maketitle

\begin{abstract}
Because of the necessity to obtain high-quality images with minimal radiation doses, such as in low-field magnetic resonance imaging, super-resolution reconstruction in medical imaging has become more popular (MRI). However, due to the complexity and high aesthetic requirements of medical imaging, image super-resolution reconstruction remains a difficult challenge. In this paper, we offer a deep learning-based strategy for reconstructing medical images from low resolutions utilizing Transformer and Generative Adversarial Networks (T-GAN). The integrated system can extract more precise texture information and focus more on important locations through global image matching after successfully inserting Transformer into the generative adversarial network for picture reconstruction. Furthermore, we weighted the combination of content loss, adversarial loss, and adversarial feature loss as the final multi-task loss function during the training of our proposed model T-GAN. In comparison to established measures like PSNR and SSIM, our suggested T-GAN achieves optimal performance and recovers more texture features in super-resolution reconstruction of MRI scanned images of the knees and belly.
\end{abstract}


\section{Introduction}
Computers have fueled our reliance on images. From the first X-rays of a tumor to the latest MRI scans, images have become integral to the practice of every field in medicine. The process of image acquisition is affected, and often limited, by many aspects, such as the equipment, environment and cost. For instance, to reduce the radiation exposed on human body, Computed Tomography (CT) is required to decrease its beam's energy, resulting in scanned images with lower spatial resolution. In the medical diagnostic process, low-quality images can affect the pathological assessment by clinical experts and auxiliary computers, among others [1-3].Calcifications, for example, are a common symptom of most breast cancers, but calcifications are small and difficult to detect. As a result, the low intensity variation between pathological tissue and healthy areas makes the diagnostic process cumbersome. In the diagnostic retinal images of fundus, many lesions cover extremely tiny areas and can be shown as microaneurysms or hemorrhages. Also, there are parts that may not clearly visible such as soft exudates, certain neointima formation, etc.[4]. Therefore, super-resolution reconstruction for medical images has become an essential role in clinical applications.

There are two main types of image super-resolution reconstruction techniques: single image super-resolution (SISR), where a high-resolution image is acquired from a single low-resolution image, and reference-based image super-resolution (RefSR), where a high-resolution image is synthesized from multiple low-resolution images. Among them, the goal of SISR often requires optimizing the mean square error between HR and SR pixels, however, the use of mean put error often leads to edge blurring due to the uncomfortable nature of super-resolution (ill-posed). The main reason for this is that the high-resolution texture of a single image is often over corrupted and a large amount of information is lost leading to unrecoverable textures[5]. While generating a large number of adversarial samples from images based on generative adversarial networks can alleviate such problems [6], the resulting hallucinations (hallucinations) and artifacts (artifacts) pose a greater challenge to image super-resolution tasks.

RefSR has been shown to be promising in providing reference (Ref) images with similar content to the LR input to recover high resolution (HR) image details [7]. A large number of RefSR methods have produced visually more pleasing results compared to SISR methods. Currently, RefSR is mainly used to make full use of the Ref image information by methods such as image aligning and "patch matching". The literature [8, 9, 10] aligns LR and Ref from different perspectives.

In [8], Landmark aligned LR and Ref by a global registration while minimizing energy; In [9], LR and Ref images need to be pre-aligned first, however non-uniform warping operation is used to enhance Ref images by matching LR and Ref feature maps to obtain super resolution; In [10], the method uses optical flow to align LR and Ref pictures at different scales and connect them to the decoder's relevant layers. However, the quality of the alignment between LR and Ref has a significant impact on the performance of these approaches. In addition, alignment methods such as optical flow require a large computational cost, making it difficult to be popularized in practical applications. On the other hand, the literature [5,11,12,13] uses a "patch matching" approach to search for suitable reference information in the Ref image to complement the information and thus obtain super resolution. In [11], the gradient features in the downsampled Ref are searched for to match the LR patch; in [12], the features in the CNN are used instead of gradient features to match the patch of Ref and LR, while the LR image is expanded using the SISR method; In [13], features in VGG are used to match the patch of Ref and LR, and super-resolution is obtained by swapping similar texture features. In [5], a texture transformer network to feature-match Ref and LR, where the low-resolution LR and reference Ref images are represented as queries and keywords in the Transformer, respectively. This setup allows the low-resolution picture LR and the reference image Ref to learn features together, i.e. deep feature correspondences that can improve accuracy in texture can be detected through an attention mechanism. However, when the reference image is less sharp, the quality of the RefSR may receive a serious impact, resulting in impaired performance of the algorithm.

In this paper, our objective is to unleash the potential of RefSR by generating reference images with more texture details through generative adversarial networks, and to discover deep feature correspondences by using the Transformer framework to perform joint feature learning between LR and reference Ref images, as shown in Figure 1, which demonstrates the effectiveness of our method on medical images. The main contributions of this paper are:
\begin{itemize}
\item Introduce a GAN framework to recover the detailed information of the original photo from the severely downsampled (low-resolution) image to obtain a high resolution image using the powerful generative power of GAN networks.
\item Introduction of Transoformer strategy to extract learnable texture features.
\end{itemize}

\section{Model Description}
In this section, we review previous classical algorithms for building single image super-resolution (SISR) for GAN-based generative adversarial networks in a single image, and classical methods for RefSR in the Transformer framework for reference-based image super-resolution, which are most relevant to our work.
\subsection{GAN in Image Reconstruction}
In image super-resolution reconstruction tasks, Generative Adversarial Networks (GANs) [6] have emerged as an effective method for enhancing the perceptual quality of upsampled images [6,14,15,16].GANs are effectively a min-max two-player game[6]. The generator G captures the data distribution, while the discriminator D continuously distinguishes whether the samples are from the training dataset or not. GANs can produce more aesthetically attractive images without supervised input using this powerful method. However, due to their intrinsic instability, the initial GANs are difficult to train. Wasserstein GANs (WGANs) use weight clipping to ensure that D is in the space of 1-Lipschit [17]; improved training of Wasserstein GANs (WGAN-GP) [18] uses gradient penalties to encourage D to learn smoother decision boundaries; and [19] proposed a weighted combination of WGANs and WGAN-GP loss terms to form a complex loss function that facilitates the model to generate. However, the above GANs are only applied on SISR while limited to image datasets with single-scale upsampling at relatively low target resolution [20]. A multi-scale GAN-enhanced SISR approach is proposed in [20], which is progressive in both architecture and training, similar to what is done in course learning, simulating the learning process from easy to hard; [21] proposes a framework for recovering its fine texture details when super-resolution at magnification factors. The framework proposes a perceptual loss function, which consists of an adversarial loss and a content loss. This adversarial loss allows the reconstructed image to be pushed towards the natural image stream shape, while constructing a discriminator network for distinguishing the super-resolution image from the original photo-realistic image.

The traditional reference learning-based image super-resolution reconstruction model (RefSR) takes the high-resolution image as the reference (Ref), so that the relevant texture is transferred to the low-resolution (LR) image. Currently, RefSR mainly makes full use of the image information of Ref by methods such as image aligning and "patch matching". However, most of the traditional methods feed all swapped features equally into the main network, neglecting to transfer high-resolution textures from the reference image using attention mechanisms, thus limiting the application of these methods in challenging situations [5,21]. Following that, study [5] proposed a new texture transformation network for image super-resolution (TTSR), in which the LR and Ref pictures are represented as queries and keywords in the transformation, respectively. A learnable texture extractor for DNNs, a relevance embedding module, a hard-attention module for texture transfer, and a soft-attention module for texture synthesis are all part of TTSR, which is tuned for the image production task. This approach supports cooperative feature learning across LR and Ref pictures, allowing attention to uncover deep feature correspondences and produce accurate texture features. The suggested texture converter can also be stacked in a cross-scale manner, allowing texture recovery at multiple magnification levels (e.g., from 1x to 4x). The method, however, relies on a high-resolution image as a reference (Ref), while in practice, a huge number of low-resolution images are frequently obtained. Therefore, we use GANS to enhance the  quality of low-resolution images from the perspective of low-resolution images as a reference, and also use the enhanced images for information complementation, and use the attention mechanism to transfer the effective features of different images to LR images to achieve RefSR reconstruction in complex environments.
\begin{figure*}[ht]
\centering
\includegraphics[width=0.8\textwidth]{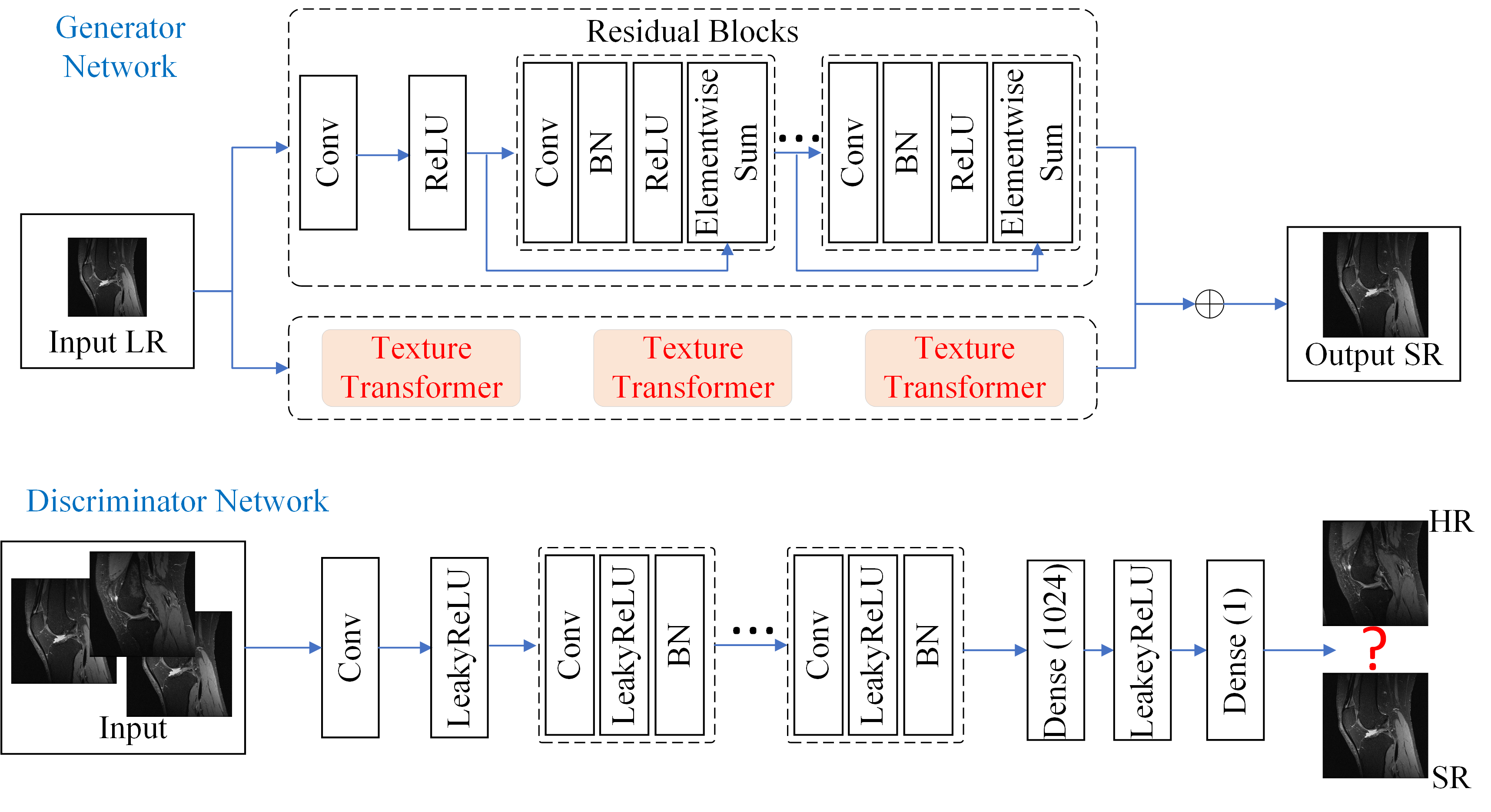}
\caption{ Schematic diagram of the model framework.}
\label{fig}
\end{figure*}
\subsection{Deep models based on Transformer and GAN}
In fact, SISR aims to learn the non-linear mapping relations between LR and HR images. In general, this non-linear mapping can be expressed as
\begin{equation}
y=\phi(k*x+n)
\end{equation}
where $\phi$ is the nonlinear compression operator, $k$ represents the convolution operation, $n$ represents the random noise, $y$ represents the degenerated LR image, and $x$ is real HR image. In general, Equation (1) can be simplified:
\begin{equation}
y=Hx
\end{equation}
where $\mathrm{H}$ is the degeneracy matrix representing the down-sampling operation. Since the conditions for the discomfort inverse issue expressed in single image super-resolution are not sufficient, $x$ can not be recovered in the simple way that
\begin{equation}
x=H^{-1}y
\end{equation}
Fortunately, deep learning based models have made a huge success in image processing fields. Many researchers have applied these deep learning models to reconstruct HR images from LR images, which actually learns an implicit mapping between LR and HR.
\begin{equation}
\hat{x}=F(y),
\end{equation}
in which $\hat{x}$ represents the reconstructed high resolution image corresponding the ground-truth image $x$. Technicality, deep neural network models minimize the optimization objective mainly by training a network $F$
\begin{equation}
\displaystyle \frac{1}{N}\sum_{i=1}^{N}(F(y_{i})-x_{i})^{2}
\end{equation}
where $N$ is the number of training samples. In general, this type of deep learning model can be represented as.
\begin{equation}
\hat{x}=F_d(\cdots F_{3}(F_{2}(F_{1}(y))))
\end{equation}
where $\mathrm{d}$ denotes the number of layers of the deep network (number of convolutional layers).

\begin{figure*}[ht]
\centering
\includegraphics[width=0.8\textwidth]{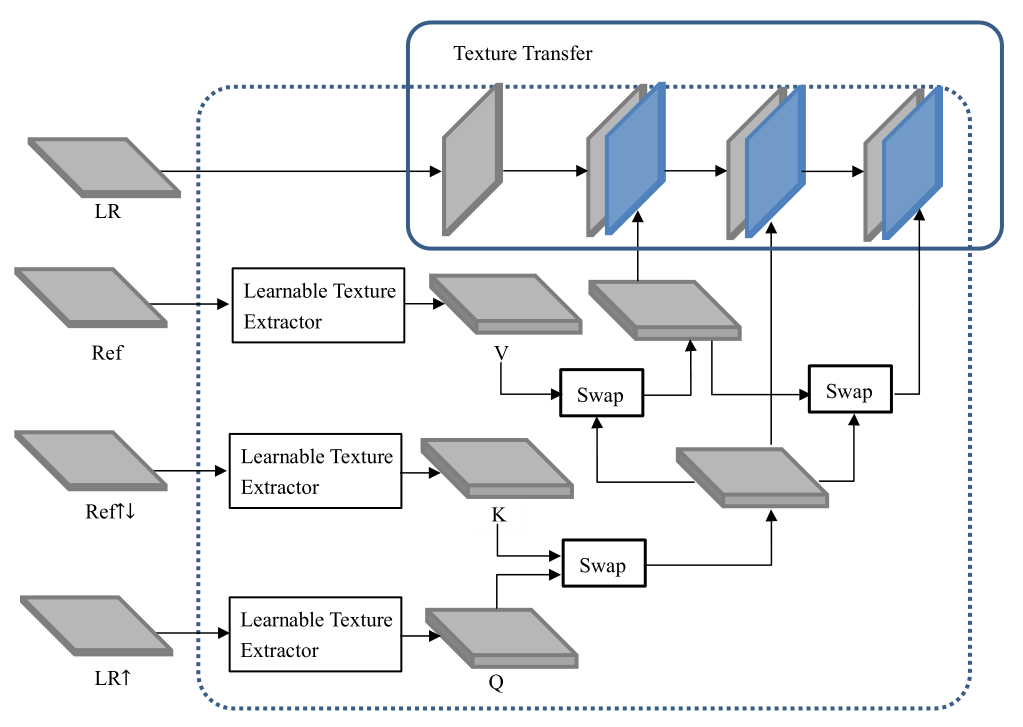}
\caption{ Schematic diagram of the texture Transformer strategy.}
\label{fig}
\end{figure*}
\subsection{Proposed Framework}
Rather than just increasing the network depth, the key goal is to increase the performance of SISR neural networks by selecting optimal internal mechanisms. The generative network and the adversarial network are the two key components of our proposed model, as depicted in Figure 1. The residual learning channel and the texture Transformer channel are two elements of the generative network.

Eventually, we intend to train a generative function G that estimates its corresponding HR image from a given LR input image. To achieve this, we propose a generative network consisting of two channels, residual learning and texture Transformer. Here we use $\theta_{G}$ to denote all parameters of the generative network as well as the bias term that is learned by optimizing a particular SR reconstruction loss $l^{SR}$. Specifically, for a given training HR image $I_{n}^{HR}$ and its corresponding LR image $I_{n}^{LR}$, the objective of the generative network is to

\begin{equation}
\displaystyle \hat{\theta}_{G}=\arg\min\frac{1}{N}l^{SR}(G_{\theta}(I_{n}^{LR}),\ I_{n}^{HR})
\end{equation}

Multiple residual learning blocks and deconvolution blocks make up the residual learning channel (as shown in Fig. 1). Because of their success in image classification, convolutional operations are now frequently utilized in deep learning, and several studies have transferred CNNs to SISR. These CNN-based SR techniques, on the other hand, rarely consider whether convolutional processes are appropriate for the SISR mechanism. The majority of them just apply CNN models to SISR from image classification tasks. The main objective of SISR is to figure out how LR and HR images are related. For the mapping relationship between LR and HR images, it can be represented by a simple linear degenerate model as follows.
\begin{equation}
y=x*k
\end{equation}
The convolution theorem states that spatial convolution can be converted to frequency-domain multiplication.
\begin{equation}
\mathcal{F}(y)=\mathcal{F}(x)\cdot \mathcal{F}(k)
\end{equation}
where $F(\cdot)$ is the Fourier transform and$\cdot$ is the corresponding element multiplication. Thus, in the Fourier domain, $x$ can be expressed as
\begin{equation}
x=\mathcal{F}^{-1}(\mathcal{F}(y)/\mathcal{F}(k))=\mathcal{F}^{-1}(1/\mathcal{F}(k))*y
\end{equation}
$\mathrm{w}\mathrm{h}\mathrm{e}\mathrm{r}\mathrm{e}\mathcal{F}^{-1}$ denotes the inverse Fourier transform and $*$ denotes the convolution operation. Thus, the true HR image can be recovered from the low-resolution image $y$ by a pseudo-inverse calculation, i.e.
$$
x=k\dagger*y
$$
where $\dagger *$ denotes the deconvolution operation.

Usually, the deconvolution kernel $k\dagger$ is hard to obtain. Therefore, we construct multiple residual learning blocks and a deconvolution block to implement the deconvolution operation. Specifically, we use a convolution kernel with a small size 3*3 and 64 feature mappings as the convolution layer followed by a batch normalization layer, while employing the ReLU function as the activation function. A residual learning mechanism is introduced (constant mapping) in order to avoid information loss and also to eliminate the gradient disappearance and gradient explosion phenomena. Finally we use the deconvolution layer (step size = 0.5) proposed by Shi et al. to improve the resolution of the input image.

For the textureTransformer channel, similar to the setup in the literature [5] (shown in Fig. 2), $\mathrm{L}\mathrm{R}, \mathrm{L}\mathrm{R}\uparrow$ and Ref denote the input image, the $ 4\times$ double-triple upsampling input image and the reference image, respectively. We apply the double triple downsampling and upsampling in turn, using the same factor $ 4\times$ on Ref to obtain $\mathrm{R}\mathrm{e}\mathrm{f}\downarrow\uparrow$ , with the domain consistent with $\mathrm{L}\mathrm{R}\uparrow$ . The texture converter accepts as input the LR features generated by Ref, $\mathrm{R}\mathrm{e}\mathrm{f}\downarrow\uparrow, \mathrm{L}\mathrm{R}\uparrow$ trunk and outputs a synthetic feature map which is further used to generate HR predictions. The texture converter consists of four components: a learnable texture extractor (LTE), a correlation embedding module (RE), a hard attention module (HA) for feature transfer, and a soft attention module (SA) for feature synthesis.

LTE mainly uses an end-to-end model to train the learning parameters such that the images of LR and Ref are able to perform joint feature learning and therefore capture more accurate texture features.LTE mainly extracts the texture features of the following three images and notates them as Q(query), K(key), V(value): $\mathrm{Q}=\mathrm{L}\mathrm{T}\mathrm{E}(\mathrm{L}\mathrm{R}$ $\uparrow), \mathrm{K}=\mathrm{L}\mathrm{T}\mathrm{E}(\mathrm{R}\mathrm{e}\mathrm{f}\downarrow\uparrow), \mathrm{V}=\mathrm{L}\mathrm{T}\mathrm{E}(\mathrm{R}\mathrm{e}\mathrm{f})$ , where LTE $()$ denotes the output of the learnable texture extractor. After extracting the texture features, the RE establishes the matching relationship between LR and Ref images by estimating the similarity between $\mathrm{Q}$ and K. First, $\mathrm{Q}$ and $\mathrm{K}$ are expanded into a number of patches (patches), which are used to compute normalized inner products to obtain the correlation between each patch; similarly, HA transfers features for the most relevant positions in each $\mathrm{Q}$ and V.As a technique to fully merge LR and Ref related information, SA employs a soft attention mechanism in which relevant texture transfers are amplified and less relevant texture transfers are avoided. In conclusion, the texture converter can effectively convert key HR texture characteristics in the reference image to LR texture features, allowing for more accurate texture production.

\subsection{Loss Function}
The perceptual loss function $l^{SR}$ definition guides the optimization direction of the generative network and is critical to the performance of the model. We use MSE to $l^{SR}$ modeling and express the perceptual loss as a weighted sum of content loss and adversarial loss components using features extracted from the textureTransformer channel, as follows.
\begin{equation}
l^{SR}=l(X^{SR})+10^{-3}l(Gen^{SR})
\end{equation}
where $l(X^{SR})$ is content loss, $l(Gen^{SR})$ is the adversarial loss.

\subsubsection{Content loss}
Traditional content loss is often based on pixel-wise MSE loss (pixel-wise MSE loss), e.g.
\begin{equation}
l_{MSE}^{SR}=\frac{1}{r^{2}WH}\sum_{x=1}^{rW}\sum_{y=1}^{rH}(I^{HR}_{x,y} - G_{\theta_{G}}(I^{LR})_{x,y})^2
\end{equation}
However this loss tends to make the model ignore the high frequency content information during training, making the solution to the problem of overly smooth textures (OTS) not ideal. Instead of relying on pixel loss, our proposed model is based on the difference between the features extracted by the texture Transformer channel, and then we define the texture-based Transformer loss as the difference between the reconstructed image $G_{\theta_{G}}(I^{LR})$ and the reference image $I^{HR}$ as the Euclidean distance between the feature representations of the reconstructed image and the reference image:
\begin{equation}
\begin{aligned}
     l_{VGG/i,j}^{SR}=  
     \frac{1}{W_{i,j}H_{i,j}} & \sum_{x=1}^{W_{i,j}}\sum_{y=1}^{H_{i,j}}(\phi_{i,j} (I^{HR})_{x,y} \\ 
     & - \phi_{i,j} (G_{\theta_{G}}(I^{LR}))_{x,y})^2
\end{aligned}
\end{equation}

\subsubsection{Adversarial Loss}

The content loss describes the difference between the reconstructed image and the reference image, while we also need to consider the loss incurred when reconstructing the image using GAN. When reconstructing an image using GAN, we need to cheat the discriminator network to obtain a more insurgent image, while generating a loss based on the probability that the discriminator produces a natural sample over all training samples defined as:
\begin{equation}
l_{Gen}^{SR}=\sum_{n=1}^{N}-\log D_{\theta_{D}}(G_{\theta_{G}}(I^{LR}))
\end{equation}
where $D_{\theta_{D}}(G_{\theta_{G}}(I^{LR}))$ denotes the reconstructed image $G_{\theta_{G}}(I^{LR})$ is the estimated probability of the natural HR image.

\section{Experiment}
In this section, the proposed model is analyzed in comparison with bicubic interpolation and some typical deep CNN-based image super-resolution reconstruction model frameworks, including EDSR [22], WDSR [23].EDSR and WDSR won the international competitions NTIRE 2017 and NTIRE 2018 image high-resolution competitions, respectively.
\subsection{Validation Indicator}
We conducted experiments on a number of benchmark medical image datasets. For a fair quantitative comparison, we use PSNR and SSIM [24] for SR framework assessment, and the evaluation indicators of PSNR and SSIM are calculated as follows.
\begin{equation}
MSE=\frac{1}{N^{2}}\sum_{i-1}^{N}\sum_{j-1}^{N}(x(i,j)-y(i,j))^{2}
\end{equation}
\begin{equation}
PSNR=10\cdot\log_{10}(\frac{MAX^{2}}{MSE})
\end{equation}
\begin{equation}
\beta_{SIM}=\frac{(2\mu_{x}\mu_{y}+c_{1})(2\sigma_{xy}+c_{2})}{(\mu^{2}_{x}+\mu_{y}^{2}+c_{1})(\sigma_{x}+\sigma_{y}+c_{2})}
\end{equation}
where $\mathrm{x}$ is generated image, $\mathrm{y}$ is ground truth image, $\mathrm{N}$ is image size, and $MAX$ is gray scale's maximum value; $\mu, \sigma$ denote the mean and variance, and $\sigma_{xy}$ denotes the covariance of the two images; two constants $c_1=(0.01-MAX)^2 $ and $c_2=(0.03-MAX)^2$  were calculated according to the SSIM convention.

\subsection{Data set and implementation details}
To validate the effectiveness of our model in real medical images, we selected separate datasets of MRI scans of the knee and abdomen datasets\footnote{http://mridata.org/about} for comparison tests. MRI imaging methods are completely different from CT images and natural images in general, and each pixel's value in MRI images has no particular physical meaning. Before training and testing, zero-mean normalization had to be applied to each MRI image (i.e., the normalization calculation is to use each value to subtract the mean and then divide by standard deviation). The low-resolution MRI image slices were obtained by averaging the $4\times 4$ pooling over the original high- resolution MRI image slices. We set $\lambda_1=5\times 10^{-2}, {\lambda_2}=5\times^{10^{-3}}$ and $\lambda \mathrm{L}_1=10^{-2}$ for training the loss function in the proposed super-resolution model and iterative optimization using the Adam [25] optimizer with $\beta_1=0.9$ and $\beta_2=0.999$, and the initial learning rate is set to $10^{-4}$.

It should be noted that we used by knee MRI images, abdominal MRI images and chest CT images[26] as the training set. To enlarge the training set, we crop the single original image into multiple small images of the same size while the downsampling factor is set to 4 to obtain low-resolution input images. Following that, the suggested depth model is trained using the obtained low-resolution dataset as well as the original high-resolution dataset.
\begin{figure}[ht]
\centering
\subfigure[Original HD image]{\includegraphics[width=4cm]{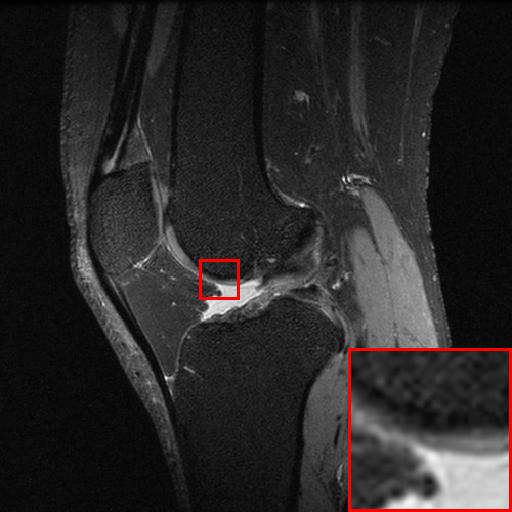}} 
\subfigure[Bicubic]{\includegraphics[width=4cm]{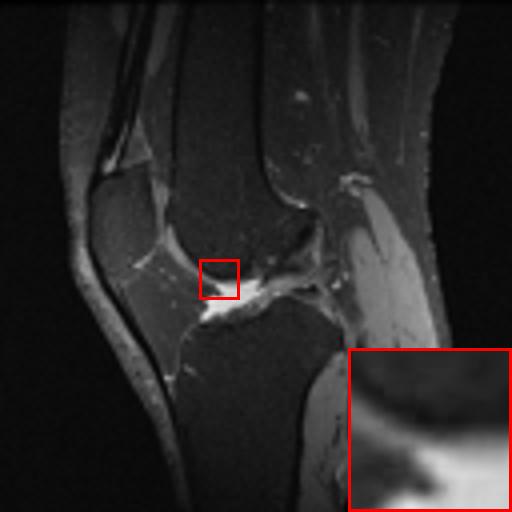}}
\\ %换行
\centering
\subfigure[EDSR]{\includegraphics[width=4cm]{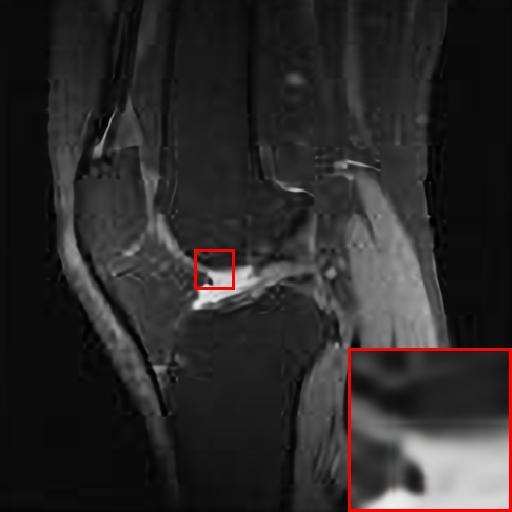}}
\subfigure[WDSR]{\includegraphics[width=4cm]{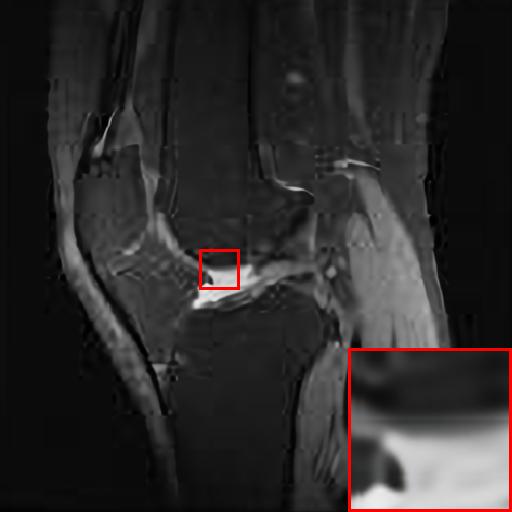}}
\\
\centering
\subfigure[T-GAN]{\includegraphics[width=4cm]{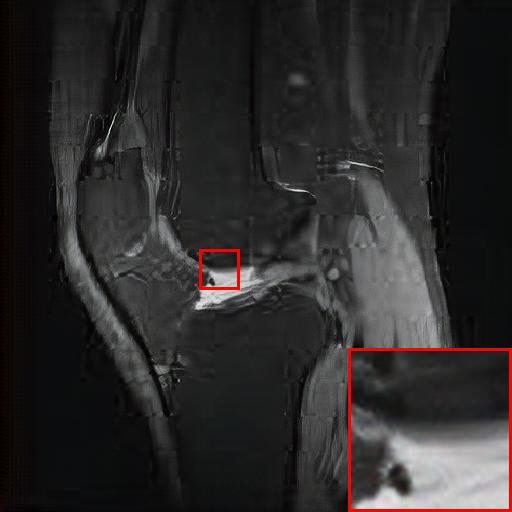}}
\caption{Reconstruction results of each algorithm for MRI images of the knee.} %图片标题
\end{figure}

\begin{figure}[ht]
\centering
\subfigure[Original HD image]{\includegraphics[width=4cm]{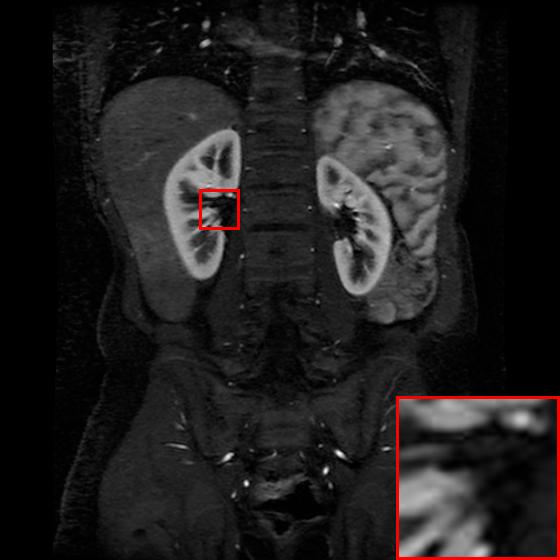}} 
\subfigure[Bicubic]{\includegraphics[width=4cm]{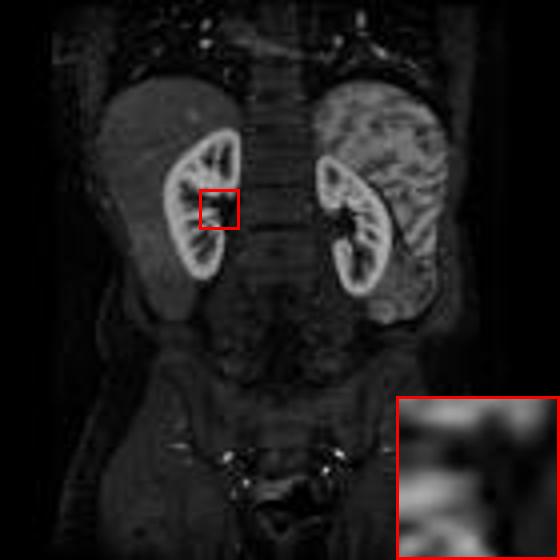}}
\\ %换行
\centering
\subfigure[EDSR]{\includegraphics[width=4cm]{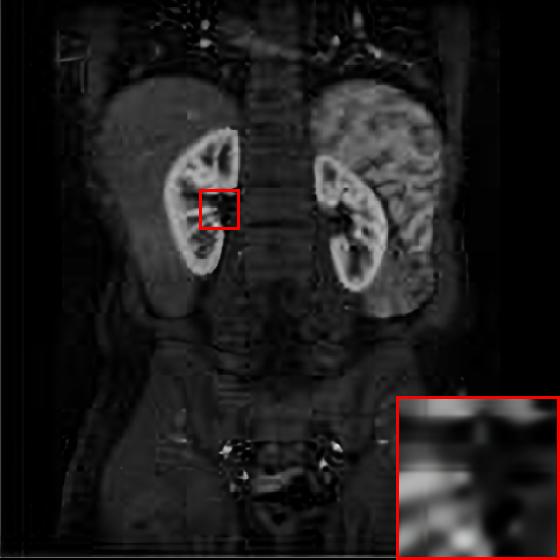}}
\subfigure[WDSR]{\includegraphics[width=4cm]{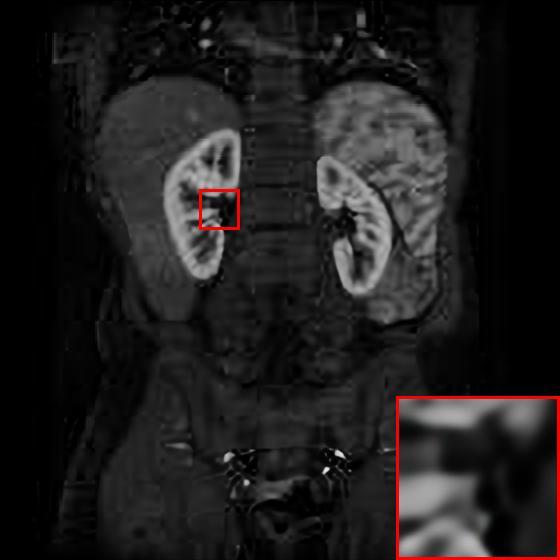}}
\\
\centering
\subfigure[T-GAN]{\includegraphics[width=4cm]{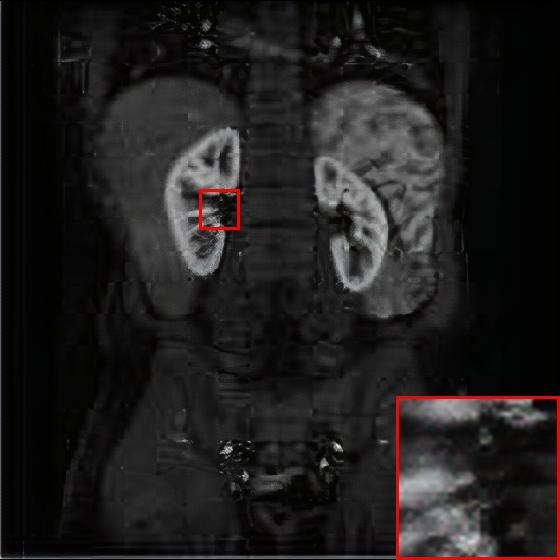}}
\caption{Reconstruction results of each algorithm for abdominal MRI images.} %图片标题
\end{figure}

\begin{figure}[ht]
\centering
\subfigure[Original HD image]{\includegraphics[width=4cm]{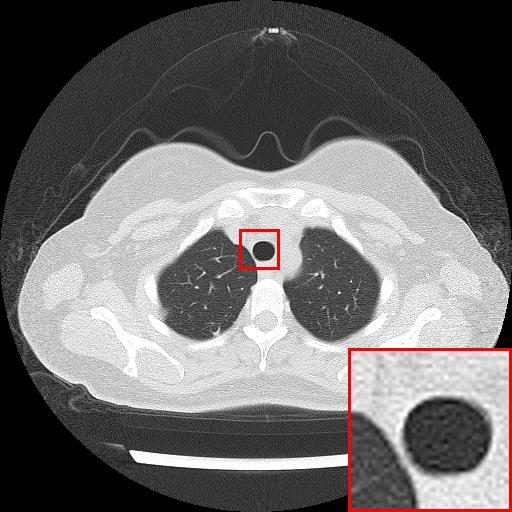}} 
\subfigure[Bicubic]{\includegraphics[width=4cm]{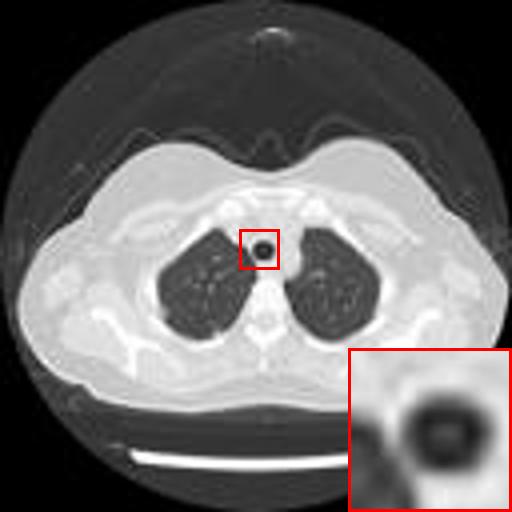}}
\\ %换行
\centering
\subfigure[EDSR]{\includegraphics[width=4cm]{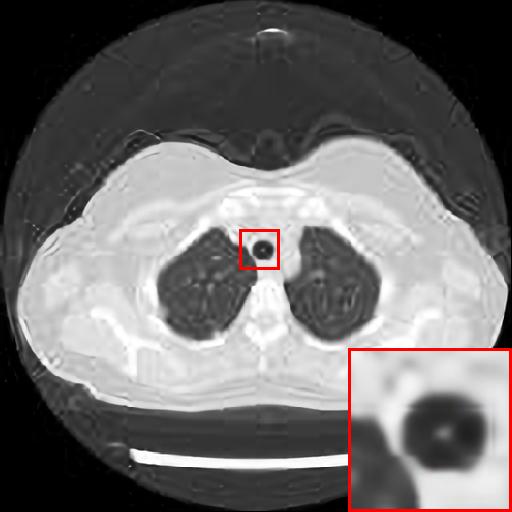}}
\subfigure[WDSR]{\includegraphics[width=4cm]{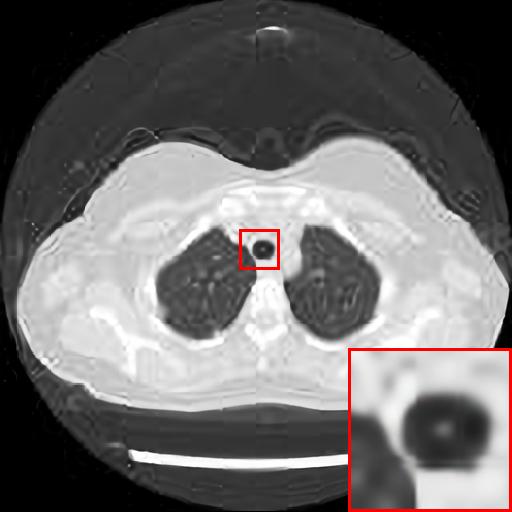}}
\\
\centering
\subfigure[T-GAN]{\includegraphics[width=4cm]{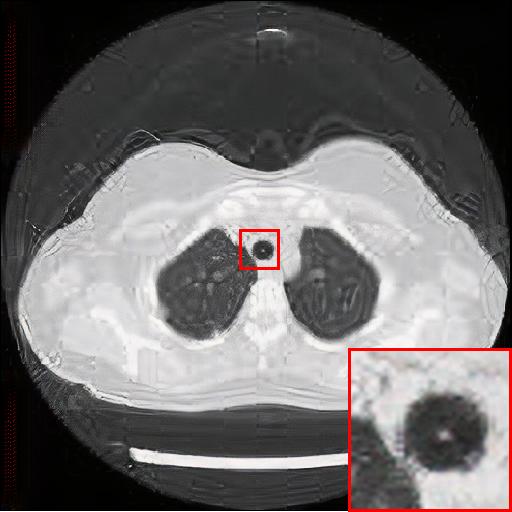}}
\caption{Reconstruction results of each algorithm for low-doze chest CT images.} %图片标题
\end{figure}

\begin{figure}[ht]
\centering
\subfigure[Original HD image]{\includegraphics[width=4cm]{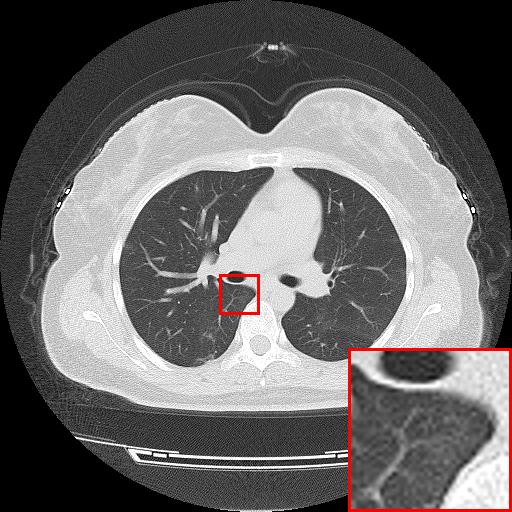}} 
\subfigure[Bicubic]{\includegraphics[width=4cm]{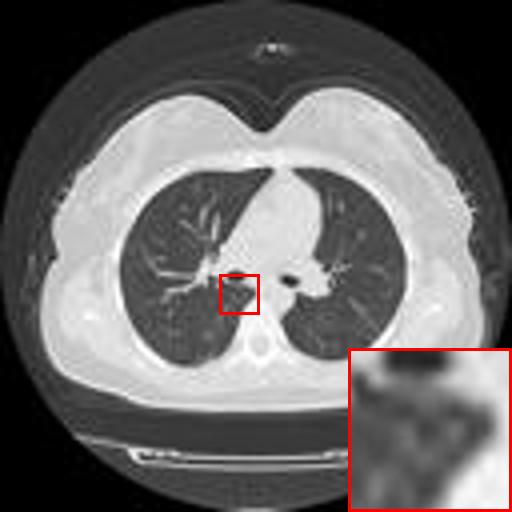}}
\\ %换行
\centering
\subfigure[EDSR]{\includegraphics[width=4cm]{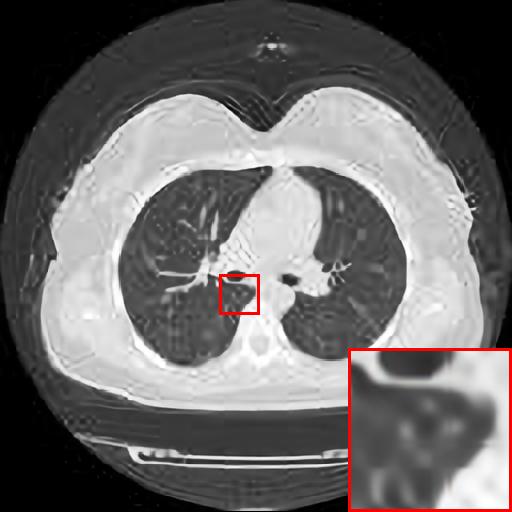}}
\subfigure[WDSR]{\includegraphics[width=4cm]{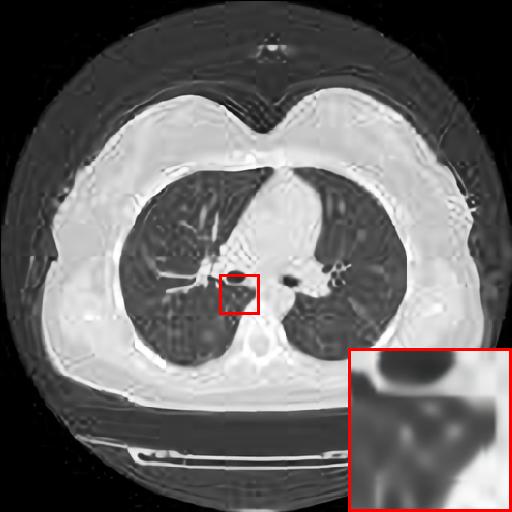}}
\\
\centering
\subfigure[T-GAN]{\includegraphics[width=4cm]{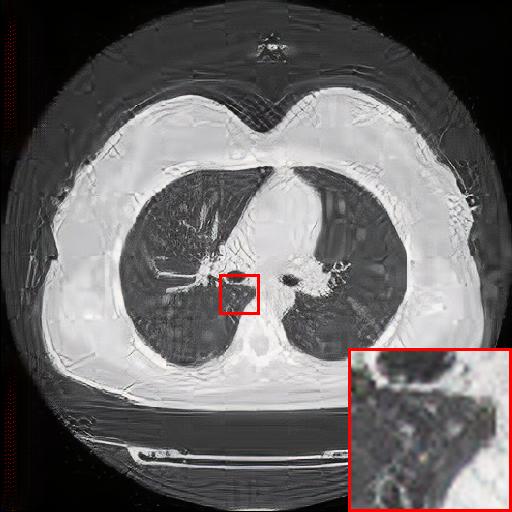}}
\caption{Reconstruction results of each algorithm for low-doze chest CT images.} %图片标题
\end{figure}

\subsection{Results and Analysis}
The PNSR/SSIM test results for the knee MRI test images for all contrast methods are shown in Table 1. The PNSR/SSIM test results for the abdominal MRI images are shown in Table 2. It's worth noting that all metrics were calculated on cropped photos in order to eliminate the impact of non-subject areas. The quantitative results show that for knee MRI images, our proposed $\mathrm{T}$-GAN model achieves the best performance on the PSNR/SSIM metrics. For abdominal MRI images, our model essentially achieves optimal performance, with individual image WDSR slightly outperforming our model. The experimental results cousin that our model is more suitable for medical image super-resolution reconstruction than the existing deep learning-based image super-segmentation models.

\begin{table*}[h]
\centering
\caption{Reconstruction results of each contrast algorithm on MRI images of the knee}
\resizebox{0.8\textwidth}{!}{
\begin{tabular}{|l|l|l|l|l|l|l|l|}
\hline
\multicolumn{2}{|l|}{}          & Image 1 & Image 2 & Image 3 & Image 4 & Image 5 & Average \\ \hline
\multirow{2}{*}{Bicubic} & PSNR & 32.92   & 31.58   & 31.88   & 28.56   & 31.21   & 31.23   \\ \cline{2-8} 
                         & SSIM & 0.9033  & 0.8802  & 0.8902  & 0.8682  & 0.8722  & 0.88282 \\ \hline
\multirow{2}{*}{EDSR}    & PSNR & 34.56   & 33.26   & 33.38   & 32.24   & 33.12   & 33.312  \\ \cline{2-8} 
                         & SSIM & 0.9405  & 0.9369  & 0.9354  & 0.9215  & 0.9324  & 0.93334 \\ \hline
\multirow{2}{*}{WDSR}    & PSNR & 34.68   & 34.12   & 33.87   & 33.18   & 34.14   & 33.998  \\ \cline{2-8} 
                         & SSIM & 0.9468  & 0.9405  & 0.9378  & 0.9289  & 0.9465  & 0.9401  \\ \hline
\multirow{2}{*}{T-GAN}   & PSNR & 35.26   & 35.03   & 35.16   & 34.17   & 34.98   & 34.92   \\ \cline{2-8} 
                         & SSIM & 0.9526  & 0.9502  & 0.9514  & 0.9453  & 0.9487  & 0.94964 \\ \hline
\end{tabular}
}
\end{table*}

\begin{table*}[h]
\centering
\caption{Reconstruction results of each contrast algorithm on abdominal MRI images}
\resizebox{0.8\textwidth}{!}{
\begin{tabular}{|l|l|l|l|l|l|l|l|l|}
\hline
\multicolumn{2}{|l|}{}          & Image 1 & Image 2 & Image 3 & Image 4 & Image 5 & Image 6 & Average \\ \hline
\multirow{2}{*}{Bicubic} & PSNR & 31.28   & 31.34   & 30.78   & 32.16   & 30.84   & 29.47   & 30.98   \\ \cline{2-9} 
                         & SSIM & 0.8842  & 0.8901  & 0.8807  & 0.8986  & 0.8872  & 0.8804  & 0.8869  \\ \hline
\multirow{2}{*}{EDSR}    & PSNR & 33.86   & 34.15   & 32.98   & 34.87   & 32.57   & 32.12   & 33.43   \\ \cline{2-9} 
                         & SSIM & 0.9026  & 0.9158  & 0.9005  & 0.9327  & 0.9124  & 0.9118  & 0.9126  \\ \hline
\multirow{2}{*}{WDSR}    & PSNR & 34.23   & 34.98   & 33.86   & 34.98   & 33.74   & 33.68   & 34.25   \\ \cline{2-9} 
                         & SSIM & 0.9242  & 0.9384  & 0.9276  & 0.9487  & 0.9305  & 0.9316  & 0.9335  \\ \hline
\multirow{2}{*}{T-GAN}   & PSNR & 35.13   & 34.62   & 34.76   & 35.13   & 34.28   & 34.19   & 34.69   \\ \cline{2-9} 
                         & SSIM & 0.9327  & 0.9318  & 0.9294  & 0.9401  & 0.9389  & 0.9391  & 0.9353  \\ \hline
\end{tabular}
}
\end{table*}

We likewise give the visualization comparison results for each comparison algorithm, as shown in Fig. 3 and Fig. 4. It can be seen that the reconstructed images based on bicubic interpolation and deep learning based EDSR and WDSR both show oversmoothing phenomenon and some loss of detail information of the images. In contrast, our T-GAN performs better for the reconstruction of detail information due to the texture Transformer structure.

\subsection{Super-resolution Reconstruction of Low-dose CT images}
Medical pictures, such as computed tomography (CT) scans, are widely used in clinical applications such as noninvasive illness detection, anatomical imaging, and treatment planning, all of which need judgment while doing CT scans. These imaging approaches, however, have some drawbacks. During CT scans, for example, radiation damage is unavoidable. Low-dose CT (LDCT) is currently the clinically recommended strategy for preventing irreversible radiation harm to the body, however it comes at the cost of getting CT pictures with low resolution or noise contamination. The spatial resolution is generally coarser than that of CT imaging in order to get images with a high signal-to-noise ratio. As a result, obtaining high-resolution scanned images with a low-dose CT scanner is now a challenge.

In this section, we selected chest CT images of COVID-19 patients in an actual hospital [26] for our experiments. The visualization results of the experiments are shown in Figure 5 and Figure 6. The experimental results show that our proposed T-GAN is also applicable to the super-resolution reconstruction of low-dose CT images, and the high-resolution images obtained by our model have more detailed information compared with the baseline algorithm.

\section{Conclusion}
In this paper, we present a super-resolution model (T-GAN) for medical pictures based on Transformer and generative adversarial network (GAN), with Tansformer approach and residual learning as two generator channels. The results suggest that our proposed T-GAN model may be employed directly for super-resolution MRI image reconstruction, and that our reconstruction methods preserve more texture information than generic image reconstruction algorithms. The findings of the experiments suggest that using the super-resolution reconstruction model to recover more picture details from clinically collected low-resolution images is possible (e.g., LDCT, low-field MRI, and MRI spectral imaging).

\vspace{12pt}
\color{red}

\end{document}